\documentclass[twocolumn]{aastex631}

\shorttitle{Planetesimal Scattering Efficiency}
\shortauthors{Stephen R. Kane \& Emma L. Miles}

\begin{document}

\title{Planetesimal Scattering Efficiency of Cold Giant Planet
  Architectures}

\author[0000-0002-7084-0529]{Stephen R. Kane}
\affiliation{Department of Earth and Planetary Sciences, University of
  California, Riverside, CA 92521, USA}
\email{skane@ucr.edu}

\author[0009-0006-9233-1481]{Emma L. Miles}
\affiliation{Department of Earth and Planetary Sciences, University of
  California, Riverside, CA 92521, USA}


\begin{abstract}

The discovery of many exoplanets has revealed an incredible diversity
of orbital architectures. These orbital configurations are
intrinsically linked to the potential for habitable environments
within the system, since the gravitational influence of the planets
governs the angular momentum distribution within the system. This
angular momentum distribution, in turn, alters the planetary orbits
and rotational obliquities. In the case of giant planets, their
gravitational influence can also produce significant redistribution of
volatiles, particularly those that lie beyond the snow line. Here, we
present the results of dynamical simulations that investigate the role
of cold giant planets in scattering material to inner terrestrial
planets. We highlight 10 exoplanetary systems with 2 or more known
giant planets beyond the snow line, and adopt a solar system analog
template that investigates the scattering of material within the range
3--8~AU. We show that increasing the eccentricity of a Jupiter analog
from its present, near-circular, value to a moderate range (0.2--0.3)
results in an order of magnitude increase in scattered material to the
inner part of the system. The inclusion of a Saturn analog to the
dynamical model produces a similar increase, highlighting the
importance of multiple giant planets beyond the snow line. However,
the addition of analogs to Uranus and Neptune can have a minor
negative effect on scattering efficiency through the transfer of
angular momentum from the inner giant planets.

\end{abstract}



\section{Introduction}
\label{intro}

A primary challenge within the subject of planetary habitability is
assessing the first, second, and third order effects on the ability of
terrestrial planets to sustain long-term temperate surface conditions
\citep{meadows2018a,kane2021e}. An important factor in determining
overall planetary habitability is the architecture and interaction of
the bodies within the system, particularly insofar as those those
interactions affect planets that lie within the Habitable Zone (HZ) of
the host star
\citep{kasting1993a,kane2012a,kopparapu2013a,kopparapu2014,kane2016c,hill2018,hill2023}. The
architecture and evolution of the solar system planetary orbits are
increasingly placed within a much broader context, as the duration and
sensitivity of exoplanet surveys continues to increase
\citep{martin2015b,horner2020b,raymond2020a,kane2021d}. Known
exoplanetary systems now number in their thousands, largely detected
using the transit and radial velocity (RV) techniques, revealing a
diversity that generally differs substantially from the solar system
\citep{ford2014,winn2015,he2019,mishra2023a}. Though observational
selection effects bias the types of planets and orbits that are
preferentially detected
\citep{ford2008a,kane2008b,zakamska2011,wittenmyer2013a}, the time
baseline of RV surveys has lasted several decades, enabling the
extension of the detection space into the outer regions of planetary
systems \citep{fischer2016}. These surveys have revealed that giant
planets beyond the snow line are relatively scarce, even for stars
similar to the Sun
\citep{wittenmyer2011a,wittenmyer2016c,wittenmyer2020b,fulton2021,rosenthal2021,bonomo2023},
highlighting the potential rarity of the outer solar system planetary
arrangement. The gravitational dominance of giant planets within a
system, and their associated locations of mean motion resonance (MMR),
can limit the formation and stability of habitable planets
\citep{raymond2006a,kopparapu2010,kane2015b,kane2020b,kane2023c}. Thus,
it is important to understand the influence that the giant planets of
the solar system, and giant planets in general, have on sculpting the
formation and evolution of terrestrial planets.

The formation of giant planets, and their consequences for the
eventual architecture and the potential habitability of HZ planets is
an active area of research
\citep{morbidelli2000,morbidelli2016b,clement2022a}, particularly as
we seek to reconcile solar system formation scenarios with the broader
exoplanetary system population
\citep{morbidelli2007b,raymond2008b,raymond2009b,kane2023a}. Within
the solar system, it is likely that significant giant planet migration
occurred, though there remains discussion regarding the extent and
timing of that migration. These migration models include the
relatively late ($\sim$700~Myr after planet formation) and chaotic
interactions of the Nice Model
\citep[e.g.,][]{gomes2005b,morbidelli2005}, and the proposed earlier
''Grand Tack'' migration of Jupiter and Saturn, which suggests that
Jupiter may have migrated inward to approach the current orbit of
Mars, before moving outward to reach its current location
\citep[e.g.,][]{walsh2011c,raymond2014a,nesvorny2018c}. Such processes
would have had a major impact on the hydration of the inner solar
system through the scattering of the volatile inventory present in the
protoplanetary disk \citep[e.g.,][]{obrien2014a,raymond2017b}. The
delivery of volatiles to the inner planets may also occur via pebble
accretion whilst the protoplanetary disk is still present
\citep{zsom2010,sato2016b,johansen2017,lambrechts2019a,johansen2021,kalyaan2023}.
An important feature of planetary systems is the location of the
``snow line'', defined as the radial distance from the center of a
protostellar disk beyond which volatiles (such as water) can
efficiently condense to form ice
\citep{ida2005,kennedy2006b,kennedy2008a,kane2011d,ciesla2014}.  The
accretion and migration of giant planets as they form beyond the snow
line can result in considerable scattering of volatile-dominated
material to the inner regions of the system
\citep{raymond2014d,raymond2017b,venturini2020b}. These scattered
volatiles may contribute to the overall volatile inventory of
terrestrial planets, both during and after their formation, providing
a foundation for the sustained presence of surface liquid water
\citep{raymond2004a,ciesla2015b,marov2018,ogihara2023}. Volatile
scattering scenarios can then support a comparative planetology
approach to studying the initial water content for Venus, Earth, and
Mars \citep{obrien2018,wilson2022c,kane2024b,lykawka2024}. The extent
of the volatile scattering also depends on various properties of the
giant planet, including their mass and orbital eccentricity, thus
subsequently affecting the potential impact scenarios for the inner
planets. The solar system formation models described above suggest
that the orbital evolution of Jupiter and Saturn likely passed through
periods of relatively high eccentricity via planet-planet interactions
and migration processes, before settling into their present
near-circular orbits \citep{pierens2014b}. On the other hand, the
eccentricity distribution derived from statistical studies of
exoplanets \citep{shen2008c,hogg2010,kane2012d,sagear2023,kane2024a}
provide a broader picture for planet formation scenarios
\citep{juric2008b,ida2013}. Given the relative dearth giant planets
beyond the snow line, it is very important to understand their role in
habitability evolution in the solar system, and in planetary systems
more generally.

In this work, we present the results of a dynamical study that
quantifies relative material scattering efficiency for different
architecture scenarios for the outer solar system. These simulations
are provided as a template for further understanding the scattering
potential of cold exoplanet giant planets in the post-gas phase of
architecture evolution. Section~\ref{science} outlines the science
motivation for this study in the context of known planetary systems
with more than one giant planet beyond the snow line.
Section~\ref{methods} describes the methodology used in the
construction of our simulations, and the various architectures
considered. Section~\ref{results} provides the detailed results of our
simulations, including the efficiency for scattering material interior
to the snow line for each of the considered architectures, and the
integrated scattering effects with respect to the orbits of the inner
planets. We discuss the implications of our results for volatile
delivery within the solar system and the relationship to exoplanetary
systems in Section~\ref{discussion}, and provide concluding remarks
and suggestions for future work in Section~\ref{conclusions}.


\section{Science Motivation}
\label{science}

\begin{figure*}
  \begin{center}
    \includegraphics[angle=270,width=17.0cm]{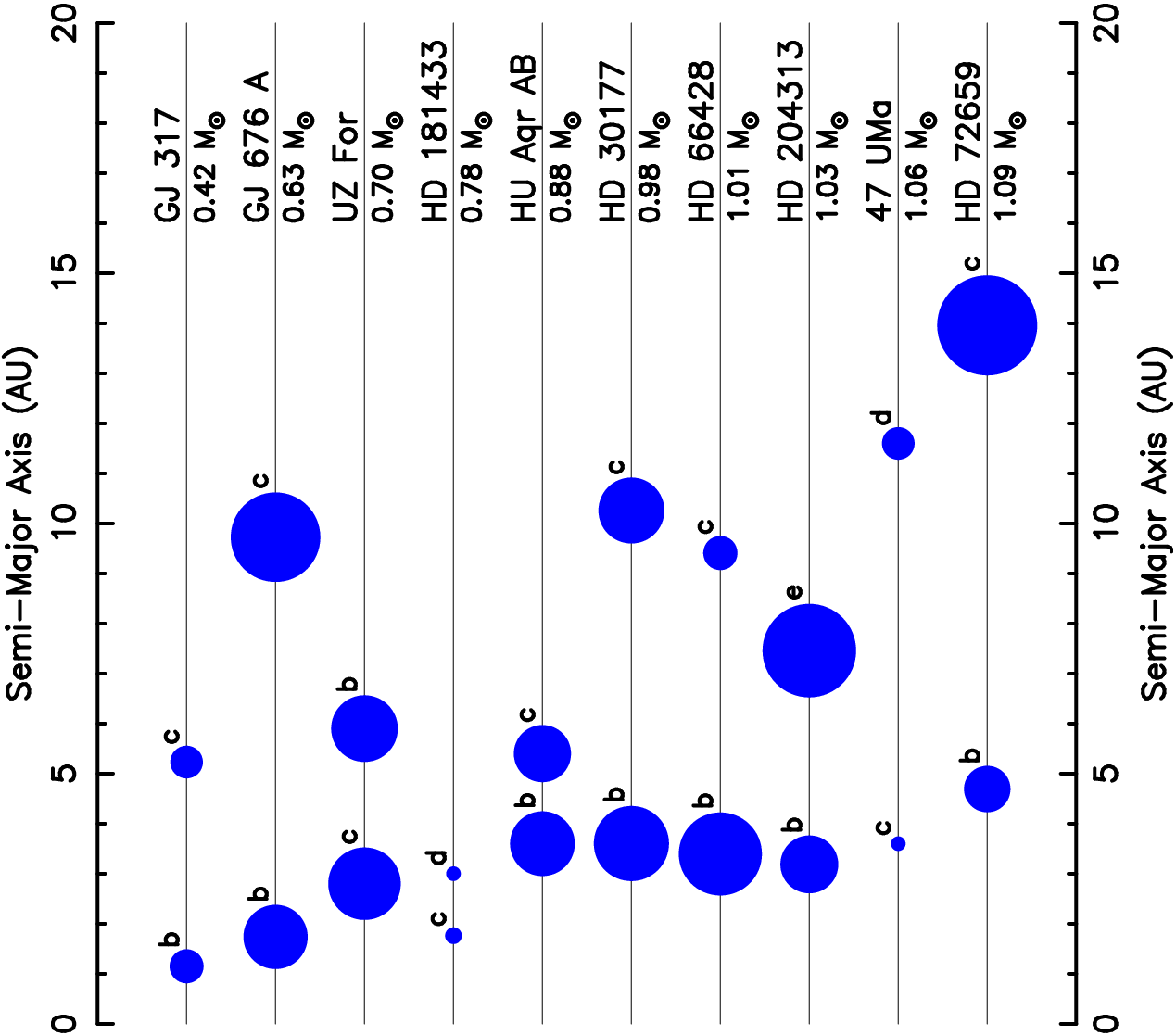}
  \end{center}
  \caption{System architectures for the 10 known planetary systems
    that have at least 2 giant planets detected beyond the snow
    line. The system architectures are shown (from top to bottom) in
    order of increasing stellar mass, which is indicated on the right
    underneath each stellar name. The size of the planets, shown in
    blue, are logarithmically proportional to the planet mass.}
  \label{fig:radii}
\end{figure*}

As described in Section~\ref{intro}, current exoplanet surveys point
toward a relative scarcity of giant planets beyond the snow line in
the majority of planetary systems \citep{wittenmyer2011a,fulton2021}.
The analysis of the cold giant exoplanet population by
\citet{kane2024a} showed that the median eccentricity for this
population is 0.23, exhibiting a far broader range of Keplerian
orbital parameters than that seen in the solar system. However, that
work did not speak to the multiplicity of such systems, only
considering the dynamical effects of a single giant planet with
various eccentricities. However, long-running RV surveys have now
probed several tens of AU around the nearest bright stars, allowing
quantitative analyses of Jupiter and Saturn analog occurrence rates to
be undertaken \citep{wittenmyer2016c,bonomo2023}.

By using the same exoplanet data as utilized by \citet{kane2024a},
extracted from the NASA Exoplanet Archive \citep{akeson2013}, we
extracted those systems for which there are two or more known giant
planets beyond the snow line. Only ten such systems are known, the
architectures of which are shown in Figure~\ref{fig:radii}. The
systems are shown in order of increasing mass of the host star,
indicated on the right side of Figure~\ref{fig:radii} underneath each
stellar name. The full inventory of these system architectures are
undoubtedly incomplete, limited by the observational bias and
measurement precision of the RV survey methodology that was used for
the bulk of the detections \citep{fischer2016}. Note, however, that HU
Aqr \citep{qian2011b} and UZ For \citep{potter2011} are circumbinary
planetary systems, where the planets were inferred via eclipse timing
variations, and the stellar mass indicated is that of the primary
star. Therefore, the dynamical environment of these two systems may
result in a divergence of the scattering efficiency of volatiles from
beyond the snow line compared with single star systems. The size of
the planets in Figure~\ref{fig:radii} (shown in blue) represent a
logarithmic proportionality to the planet mass. The planet masses
within the ten systems range from 0.54~$M_J$ (HD~181433~d and
47~Uma~c) to 18.81~$M_J$ (HD~72659~c). The letter designation is also
shown next to each of the planets.

\begin{figure}
  \includegraphics[angle=270,width=8.5cm]{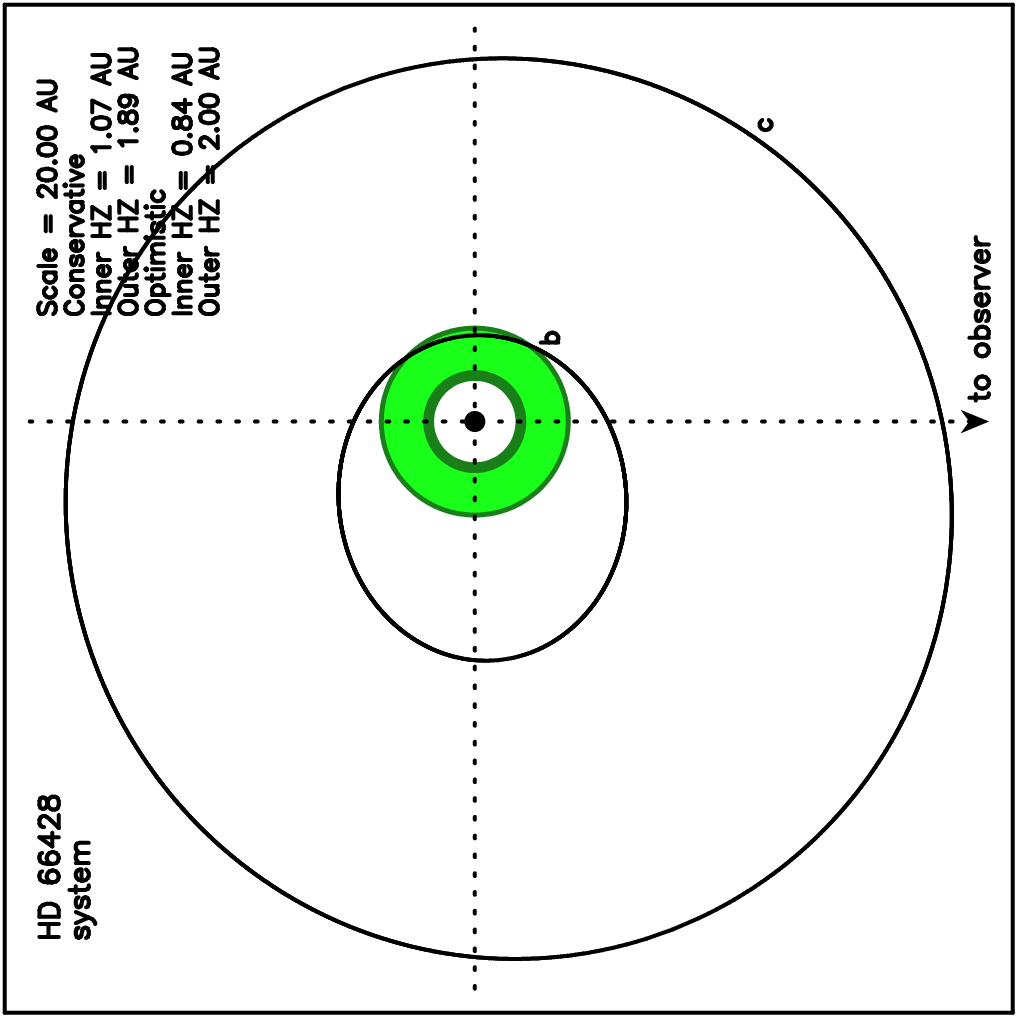}
  \caption{A top-down view of the HZ and planetary orbits in the
    HD~66428 system, where the orbits are labeled by planet
    designation. The extent of the HZ is shown in green, where light
    green and dark green indicate the CHZ and OHZ, respectively. The
    scale of the figure is 20~AU along each side.}
  \label{fig:morbit}
\end{figure}

The eccentricity distribution of the ten system sample is similarly
diverse, with eccentricities ranging from circular to 0.51
(HU~Aqr~AB~c). HD~66428 is an example of a system in which
eccentricities beyond the snow line far exceed those seen in our solar
system. Shown in Figure~\ref{fig:morbit} is a top-down view of the
system architecture of HD~66428, where the Keplerian planetary orbits
use the results provided by \citet{feng2022a}. The scale of the figure
is 20~AU along each side. Also shown is the extent of the HZ,
including the conservative HZ (CHZ) and optimistic HZ (OHZ), shown in
light and dark green, respectively. The CHZ adopts the traditional
runaway and maximum greenhouse boundaries, whilst the OHZ is based
upon assumptions regarding the prevalence of surface liquid water for
Venus and Mars
\citep{kasting1993a,kopparapu2013a,kopparapu2014,kane2016c}.

The presence of Jupiter and Saturn has placed our solar system within
a (thus far) relatively rare category of planetary system
architectures that contain significant sources of gravitational
perturbations beyond the snow line. However, the range of masses and
eccentricities of the systems portrayed in Figure~\ref{fig:radii}
suggests a complexity of interactions between the planets and/or the
disk during formation. Although the vast majority of planetesimal will
have merged with the giant planets during formation or been subject to
drag and migration during the gas phase \citep{deienno2018}, the
resulting architectures of these systems motivates a study of the
scattering potential for remaining material after the protoplanetary
gas disk dispersal. In the subsequent analysis, we adopt various
orbital configurations for the solar system planets as a template for
exploring such planetesimal scattering efficiencies as a template for
exoplanetary scenarios.


\section{Simulation Methodology}
\label{methods}

\begin{deluxetable*}{lrrrrrrrrrrrrrrrr}
  \tablecolumns{17}
  \tablewidth{0pc}
  \tablecaption{\label{tab:mmr} MMR locations for solar system outer
    planets.}
  \tablehead{
    \colhead{Planet} &
    \colhead{3:1} &
    \colhead{5:2} &
    \colhead{7:3} &
    \colhead{2:1} &
    \colhead{7:4} &
    \colhead{5:3} &
    \colhead{3:2} &
    \colhead{7:5} &
    \colhead{5:7} &
    \colhead{2:3} &
    \colhead{3:5} &
    \colhead{4:7} &
    \colhead{1:2} &
    \colhead{3:7} &
    \colhead{2:5} &
    \colhead{1:3}
  }
  \startdata
Jupiter &  2.50 &  2.82 &  2.96 &  3.28 &  3.58 &  3.70 &  3.97 &  4.16 &  6.51 &  6.81 &  7.31 &  7.55 &  8.25 &  9.15 &  9.58 & 10.82 \\
Saturn  &  4.61 &  5.20 &  5.45 &  6.04 &  6.60 &  6.82 &  7.31 &  7.66 & 11.99 & 12.55 & 13.47 & 13.91 & 15.21 & 16.85 & 17.65 & 19.93 \\
Uranus  &  9.24 & 10.43 & 10.92 & 12.10 & 13.23 & 13.67 & 14.66 & 15.35 & 24.04 & 25.17 & 27.00 & 27.90 & 30.49 & 33.79 & 35.39 & 39.96 \\
Neptune & 14.48 & 16.35 & 17.12 & 18.97 & 20.73 & 21.42 & 22.98 & 24.06 & 37.68 & 39.46 & 42.33 & 43.73 & 47.80 & 52.97 & 55.46 & 62.63 \\
  \enddata
\end{deluxetable*}

The perturbative influence of giant planets on material beyond the
snow line may be assessed through a suite of dynamical simulations
that calculate the change in orbits for that material. To carry out
this experiment, we utilized the Mercury Integrator Package
\citep{chambers1999} with a hybrid symplectic/Bulirsch-Stoer
integrator and a Jacobi coordinate system
\citep{wisdom1991,wisdom2006b}. The methodology described here follows
the prescription that has previously been successfully applied
\citep{kane2014b,kane2019c,kane2021a,kane2023c,kane2024a}. Of
particular importance to our analysis are the MMR locations for the
solar system giant planets. For convenience, and as a reference to
subsequent discussion, we provide the primary MMR locations in
Table~\ref{tab:mmr}. Specifically, the shown MMR locations are those
produced by the solar system giant planet orbits that affect the
particles that are included in our simulations.

\begin{deluxetable}{lll}
  \tablecolumns{6}
  \tablewidth{0pc}
  \tablecaption{\label{tab:sims} Simulation summary (see
    Figure~\ref{fig:sims}).}
  \tablehead{
    \colhead{Sim} &
    \colhead{Time} &
    \colhead{Description} \\
    \colhead{} &
    \colhead{(years)} &
    \colhead{}
  }
  \startdata
1 & $10^5$ & Jupiter analog ($e = 0.049$) \\
2 & $10^5$ & Eccentric Jupiter ($e = 0.23$) \\
3 & $10^6$ & Jupiter analog ($e = 0.049$) \\
4 & $10^6$ & Eccentric Jupiter ($e = 0.23$) \\
5 & $10^6$ & Eccentric Jupiter ($e = 0.1$) \\
6 & $10^6$ & Eccentric Jupiter ($e = 0.3$) \\
7 & $10^6$ & Jupiter--Saturn analog \\
8 & $10^6$ & Jupiter--Saturn--Uranus--Neptune analog \\
  \enddata
\end{deluxetable}

\begin{figure}
  \begin{center}
     \includegraphics[angle=270,width=8.5cm]{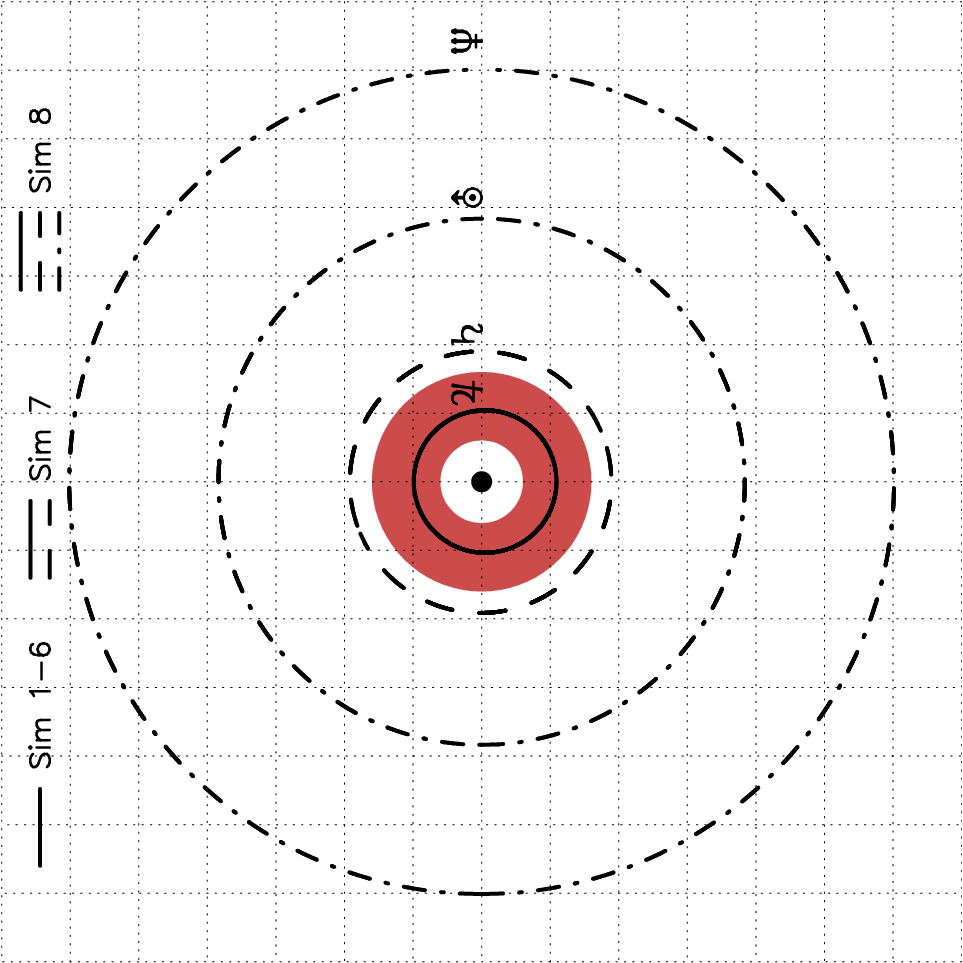}
  \end{center}
  \caption{Top-down view of the planetary systems described in
    Table~\ref{tab:sims} and the orbits that are included in each
    simulation. The grid scale is 5~AU and the red shaded region
    indicates the semi-major axis range over which particles are
    injected for all simulations. Simulations 1--6 include a Jupiter
    analog with variable eccentricities and simulation
    durations. Simulation 7 adds a Saturn analog and simulation 8
    further adds Uranus and Neptune analogs.}
  \label{fig:sims}
\end{figure}

We conducted 8 primary simulation suites that explore a range of
eccentric Jupiter cases, as well as solar system analog scenarios. The
simulation scenarios are summarized in Table~\ref{tab:sims} and the
accompanying Figure~\ref{fig:sims}. In all cases, we adopt a solar
mass for the host star, and the planetary masses are set to those of
the corresponding solar system giant planets. The orbital properties
of the giant planets were extracted from the Jet Propulsion Laboratory
(JPL) Planetary and Lunar Ephemerides DE440 and DE441
\citep{park2021}. The simulations within each suite explored a
location range of 3.0--8.0~AU in steps of 0.01~AU, resulting in
$\sim$500 simulations for each of the simulation suites. The
simulations at each semi-major axis step were run for $10^5$~years for
simulations 1 and 2, and $10^6$~years for simulations 3--8, with a
time step of 10 days. For every semi-major axis location, 100
particles were injected into circular orbits at equally spaced
starting locations. The mass of the particles was chosen to be
arbitrarily small, with a mass of $10^{-6}$ Earth masses. However, we
also tested particle masses as low as $10^{-12}$ Earth masses, which
provided indistinguishable simulation outcomes. At the conclusion of
the simulation for a given semi-major axis location, the "scattering
efficiency" of the simulation architecture for the specified location
is calculated from the percentage of particles that, at any point
during the simulation, are scattered into an orbit whose perihelion
location, $q$, is interior to a given semi-major axis threshold. The
scattering thresholds used are the orbits of Mercury ($q = 0.387$~AU),
Venus ($q = 0.723$~AU), Earth ($q = 1.000$~AU), Mars ($q = 1.524$~AU),
and the snow line ($q = a_\mathrm{ice} = 2.7$~AU), derived from
\citet{ida2005}.

Simulations 1--6 include only Jupiter, using a variety of orbital
eccentricity scenarios, to isolate the scattering efficiency of
Jupiter. Simulations 1 and 2 are the Jupiter analog ($e = 0.049$) and
eccentric Jupiter ($e = 0.23$) scenarios described by
\citet{kane2024a}. Simulations 3 and 4 extend the integration time of
1 and 2 from $10^5$ years to $10^6$ years. The eccentricity of $e =
0.23$ for simulations 2 and 4 are derived from the median eccentricity
of known exoplanets beyond the snow line with masses greater than
0.3~$M_J$ \citep{kane2024a}. Simulations 5 and 6 test two further
eccentricity scenarios for Jupiter; $e = 0.1$ and $e =
0.3$. Simulation 7 uses both current Jupiter and and Saturn
analogs. Simulation 8 includes the full suite of solar system giant
planets; Jupiter, Saturn, Uranus, and Neptune. Simulations 7 and 8
both adopt the present orbital properties for the solar system giant
planets.


\section{Dynamical Results}
\label{results}

We present the simulations results in three main categories: Jupiter
for various timescale and eccentricity configurations (simulations
1--6), Jupiter and Saturn (simulation 7), and all four giant planets
(simulation 8), as described in Table~\ref{tab:sims}.


\subsection{Jupiter's Orbital Eccentricity}
\label{ecc}

\begin{figure*}
  \begin{center}
    \includegraphics[angle=270,width=17.0cm]{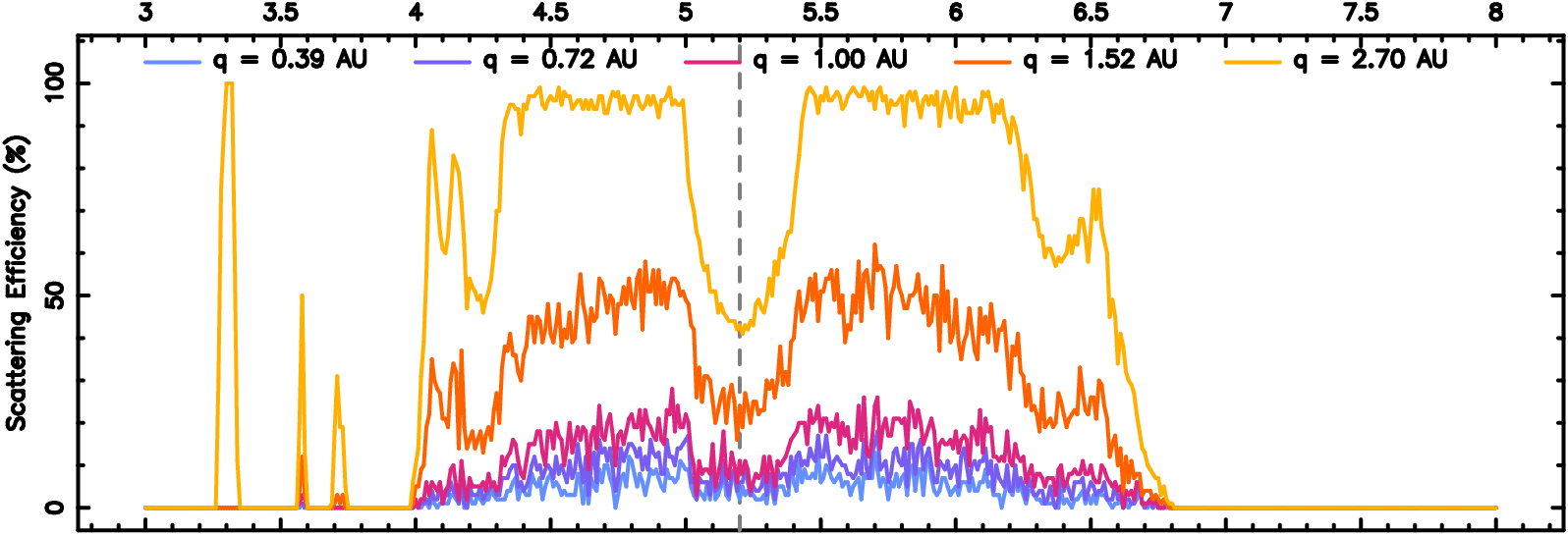} \\
    \includegraphics[angle=270,width=17.0cm]{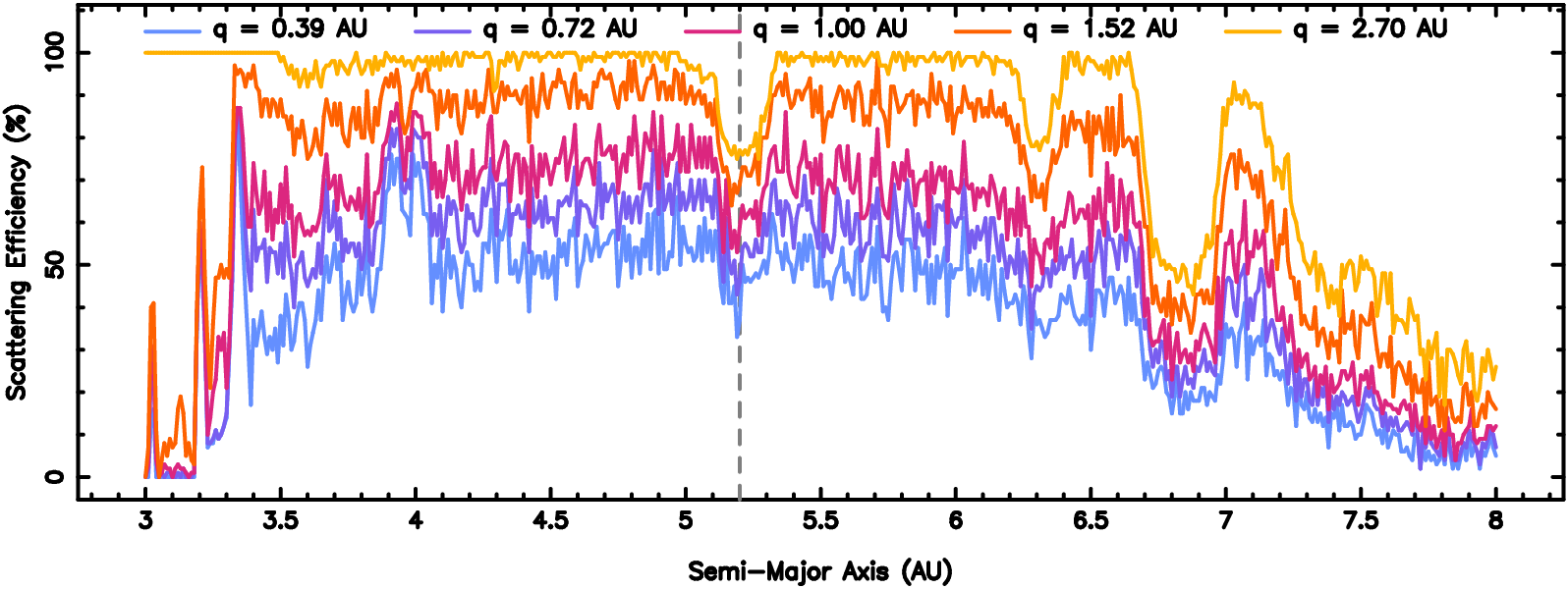}
  \end{center}
  \caption{Results for particle injection simulations 1 and 2, that
    test the scattering efficiency of a Jupiter analog ($e = 0.05$;
    top panel) compared with an eccentric Jupiter ($e = 0.23$; bottom
    panel), where each location is integrated for $10^5$~years. For
    each panel, results are plotted for the percentage of particles
    that achieve periastron passages interior to the snow line and the
    orbits of Mars, Earth, Venus, and Mercury. The gray vertical
    dashed line indicates the semi-major axis of Jupiter.}
  \label{fig:sim12}
\end{figure*}

Jupiter's dominant planetary mass within the solar system and
proximity to the snow line produces a powerful configuration for
scattering material to the interior parts of the
system. \citet{kane2024a} performed a suite of $10^5$~year dynamical
simulations that compared a Jupiter analog to an eccentric Jupiter
scenario, corresponding to simulations 1 and 2 in
Table~\ref{tab:sims}. We repeat those simulations here to further
incorporate the additional scattering thresholds described in
Section~\ref{methods}, the results of which are shown in
Figure~\ref{fig:sim12}. The scattering efficiency for the various
semi-major axis threshold are indicated by the colored lines, and the
vertical dashed line shows the semi-major axis location for
Jupiter. As noted by \citet{kane2024a}, there is a significant drop in
scattering efficiency for the Jupiter analog case (simulation 1) near
the semi-major axis of Jupiter due to the effect of Trojan particles
\citep{levison1997a,morbidelli2005,nesvorny2013a,bottke2023b}, an
effect that is less noticeable for the eccentric Jupiter case
(simulation 2). Note that, in this context, the term ``Trojan
particles'' is being used to refer to those particles whose initial
semi-major axis is at or near the Jupiter semi-major axis and not
necessarily located at the L4/L5 Lagrange points. Also of note are the
2:1 ($\sim$3.3~AU), 7:4 ($\sim$3.6~AU), and 5:3 ($\sim$3.7~AU) MMR
locations (see Table~\ref{tab:mmr}), that are generally effective at
scattering material interior to the snow line, but not to the inner
planets. The eccentric Jupiter case, shown in the bottom panel of
Figure~\ref{fig:sim12}, is far more efficient at scattering material
to the inner solar system. However, note that there is a large drop in
scattering efficiency at $\sim$6.8~AU, corresponding to the 2:3
MMR. Particles in or near the 2:3 MMR location have the potential to
avoid gravitational perturbing interactions with Jupiter, similar to
the effect of the Trojan particles described above, and the 2:3 MMR of
Neptune with Pluto \citep{williams1971e}.

\begin{figure*}
  \begin{center}
    \includegraphics[angle=270,width=17.0cm]{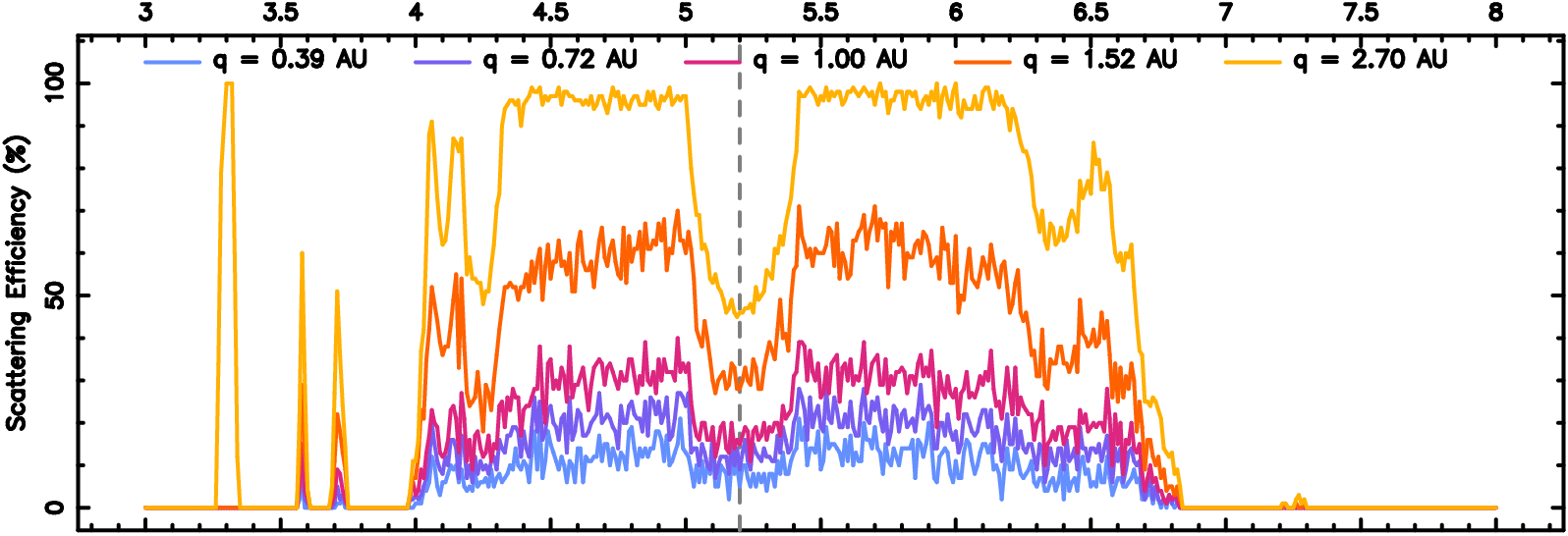} \\
    \includegraphics[angle=270,width=17.0cm]{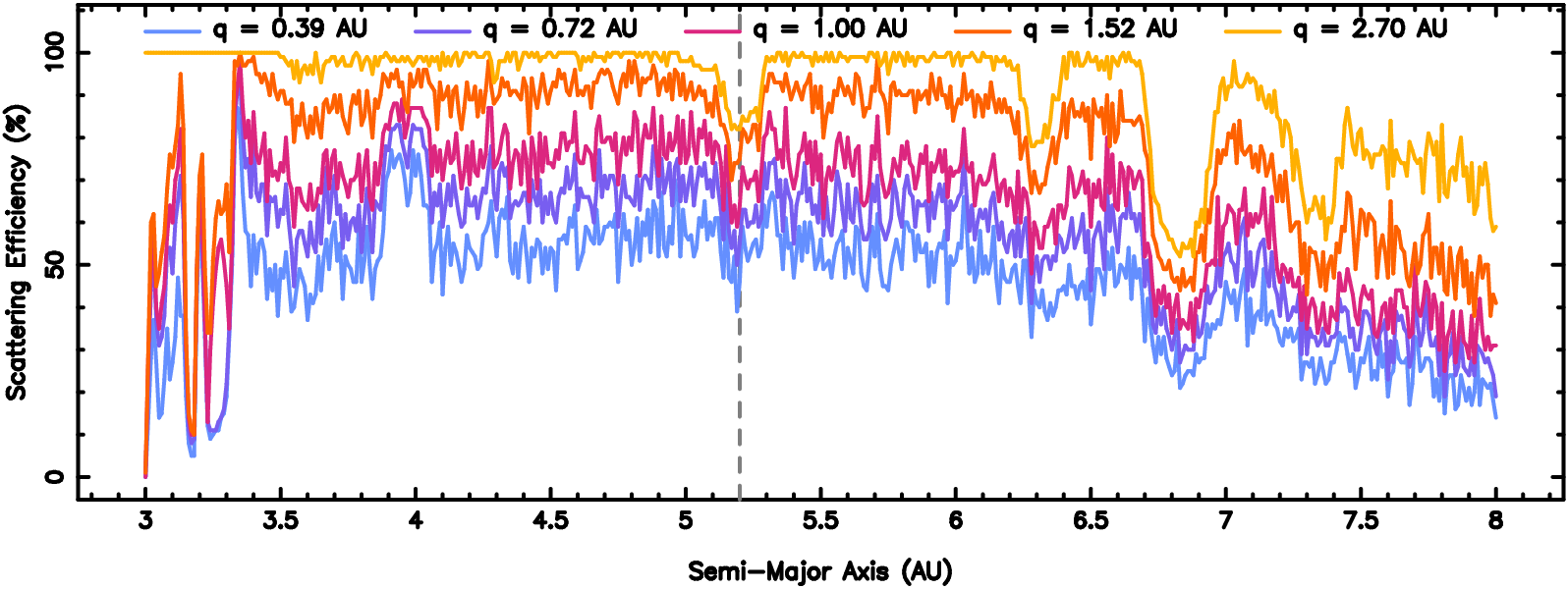}
  \end{center}
  \caption{Results for particle injection simulations 3 and 4, that
    test the scattering efficiency of a Jupiter analog ($e = 0.05$;
    top panel) compared with an eccentric Jupiter ($e = 0.23$; bottom
    panel), where each location is integrated for $10^6$~years. For
    each panel, results are plotted for the percentage of particles
    that achieve periastron passages interior to the snow line and the
    orbits of Mars, Earth, Venus, and Mercury. The gray vertical
    dashed line indicates the semi-major axis of Jupiter.}
  \label{fig:sim34}
\end{figure*}

Simulations 3 and 4 adopt the same orbital architectures as simulations
1 and 2, but extend the integration time to $10^6$~years. The results
from simulations 3 and 4 are shown in Figure~\ref{fig:sim34}. The
order of magnitude increase in the integration time produces a slight
increase in the overall scattering efficiency in both cases. One of
the most notable changes is the outer semi-major axis regions for
simulation 4, where the longer integration time allows the scattering
potential of the planet to be more fully realized.

\begin{figure*}
  \begin{center}
    \includegraphics[angle=270,width=17.0cm]{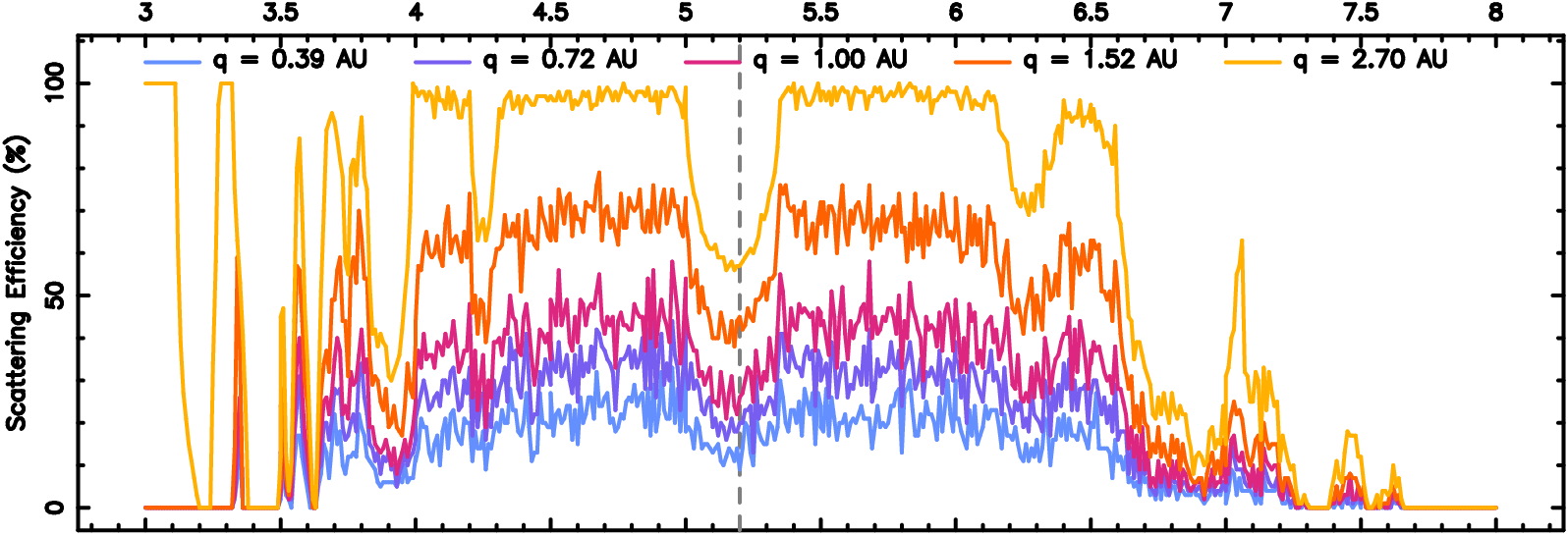} \\
    \includegraphics[angle=270,width=17.0cm]{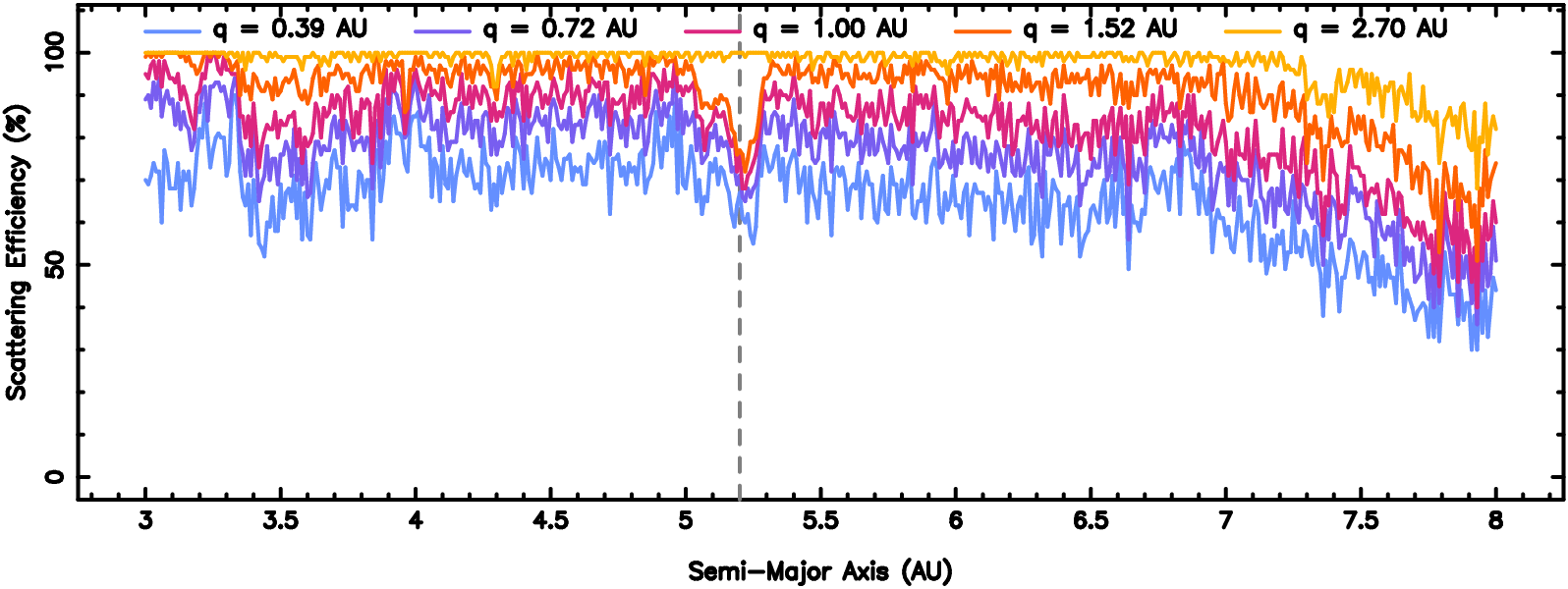}
  \end{center}
  \caption{Results for particle injection simulations 5 and 6, that
    test the scattering efficiency of two different eccentric Jupiter
    scenarios: $e = 0.1$ (top panel) and $e = 0.3$ (bottom panel),
    where each location is integrated for $10^6$~years. For each
    panel, results are plotted for the percentage of particles that
    achieve periastron passages interior to the snow line and the
    orbits of Mars, Earth, Venus, and Mercury. The gray vertical
    dashed line indicates the semi-major axis of Jupiter.}
  \label{fig:sim56}
\end{figure*}

The results of simulations 5 and 6, shown in Figure~\ref{fig:sim56},
provide the scattering efficiency outcomes from two further
eccentricities of Jupiter. In the top panel, an eccentricity of $e =
0.1$ was adopted, and the bottom panel shows the $e = 0.3$ case. These
eccentricity values were selected based on on the 1$\sigma$ RMS scatter
of eccentricities for giant planets beyond the snow line, described by
\citet{kane2024a}. The $e = 0.1$ case, shown in the top panel of
Figure~\ref{fig:sim56}, starts to noticeably exhibit significant
scattering increases compared with simulations 1 and 3, which both
adopt the present Jupiter eccentricity. Specifically, the primary
increases occur interior to the 3:2 MMR ($\sim$4.0~AU) and exterior to
the 2:3 MMR ($\sim$6.8~AU) locations, which effectively demarcate the
boundaries of primary scattering efficiency for the Jupiter analog
scenarios. The $e = 0.3$ case, shown in the bottom panel of
Figure~\ref{fig:sim56}, results in near 100\% of material being
scattered interior to the snow line, and more than 50\% of material
being scattered into Mercury-crossing orbits, for the entire
semi-major axis range considered. Such high scattering may be
unsurprising given that the perihelion and aphelion of Jupiter in this
scenario are 3.64~AU and 6.76~AU, respectively. Therefore, it is
expected that Jupiter eccentricities beyond $e = 0.3$ asymptotically
approach the maximum scattering potential for the entire semi-major
axis range explored.


\subsection{Jupiter and Saturn}
\label{sat}

\begin{figure*}
  \begin{center}
    \includegraphics[angle=270,width=17.0cm]{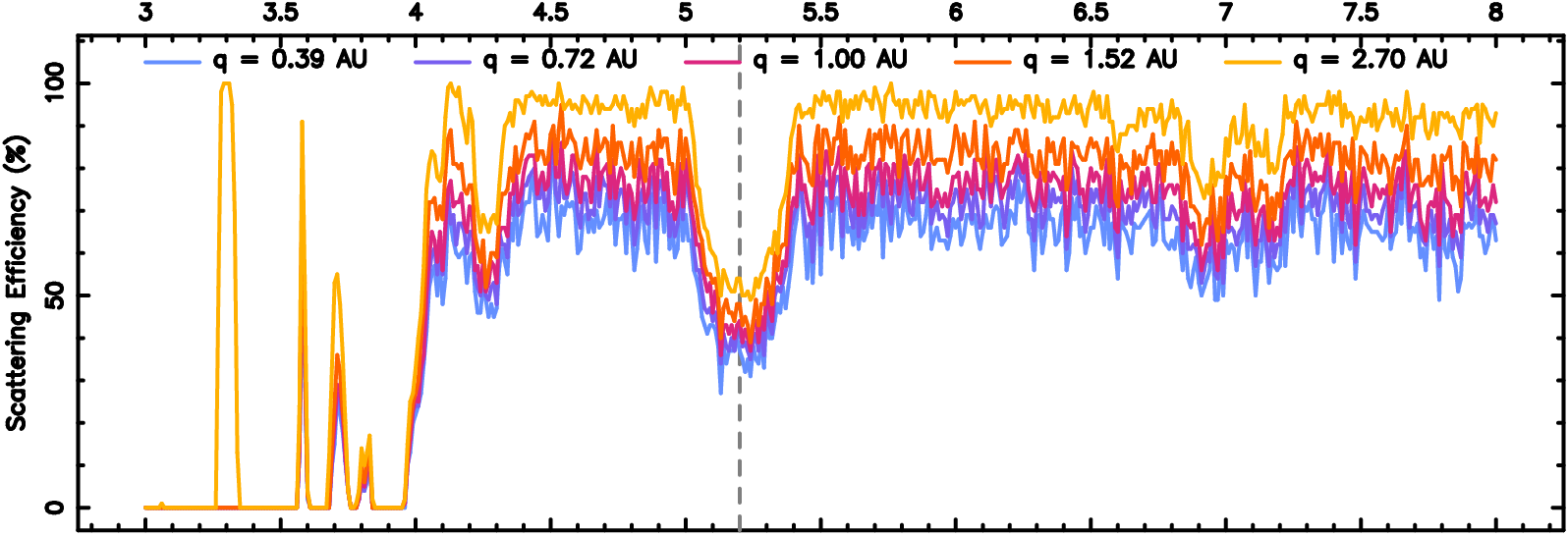} \\
    \includegraphics[angle=270,width=17.0cm]{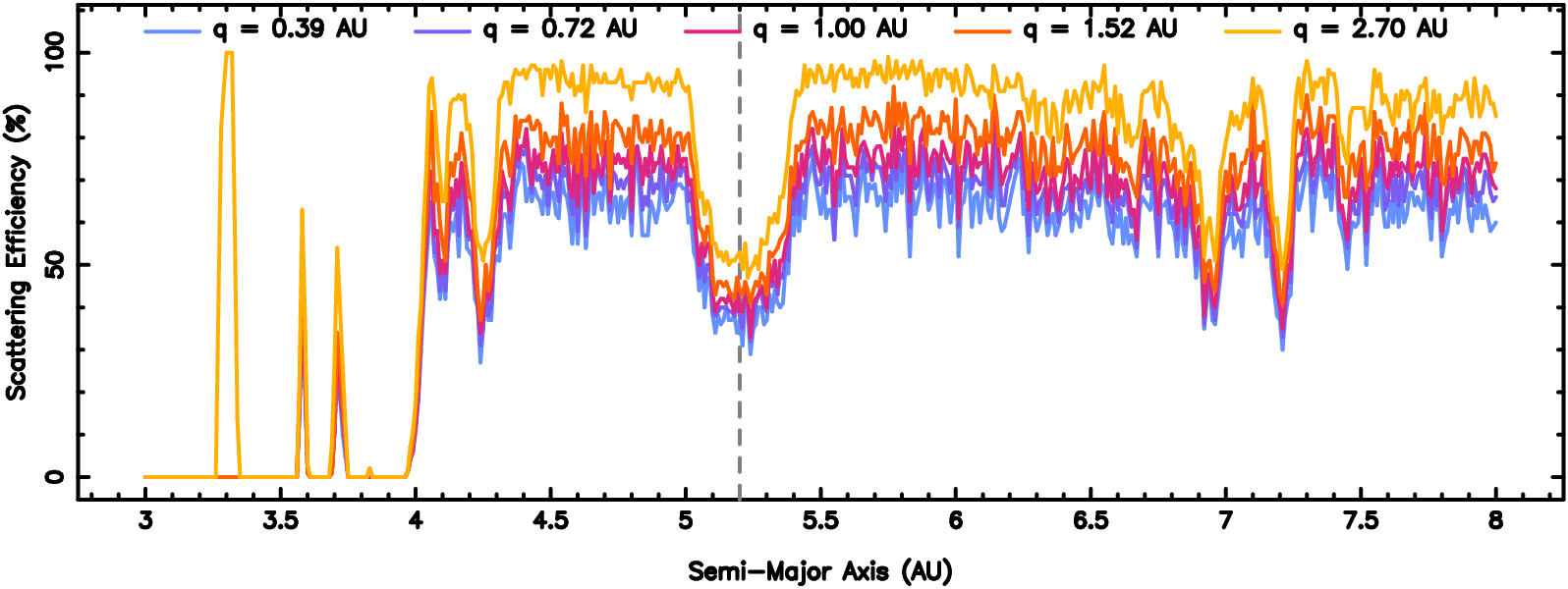}
  \end{center}
  \caption{Results for particle injection simulations 7 and 8, that
    test the scattering efficiency of a Jupiter and Saturn analog (top
    panel) and the complete solar system giant planet suite of
    Jupiter, Saturn, Uranus, and Neptune (bottom panel), where each
    location is integrated for $10^6$~years. For each panel, results
    are plotted for the percentage of particles that achieve
    periastron passages interior to the snow line and the orbits of
    Mars, Earth, Venus, and Mercury. The gray vertical dashed line
    indicates the semi-major axis of Jupiter.}
  \label{fig:sim78}
\end{figure*}

Simulation 7 adopts the current orbits for both Jupiter and
Saturn. The inclusion of Saturn allows an assessment of Saturn's
contribution to the scattering efficiency, particularly in comparison
to simulation 3, which includes a Jupiter analog for the same
$10^6$~year time period. The addition of Saturn also increases the
relevance of the scattering results to the relatively rare scenario of
at least two giant planets beyond the snow line (see
Figure~\ref{fig:radii}). The results for simulation 7 are shown in the
top panel of Figure~\ref{fig:sim78}. There are several features to
note regarding these simulation results. The relative lack of
scattering interior to the 3:2 MMR location ($\sim$4.0~AU) is similar
to that observed for the Jupiter analog case (simulations 1 and 3). As
one may expect, the presence of Saturn clearly creates a strong
perturbing influence for the semi-major axis range exterior to the 3:2
MMR location that extends to 8~AU and beyond. Indeed, the scattering
of material, even as far as Mercury-crossing orbits, is considerably
more efficient than the Jupiter analog case. A primary cause for this
increased scattering efficiency is the introduction of angular
momentum exchange between Jupiter and Saturn that produces
eccentricity ranges of 0.0454--0.0612 and 0.0151--0.0684 for Jupiter
and Saturn, respectively \citep{perminov2020,mikkola2022}. Indeed, the
secular eccentricity oscillations of Jupiter and Saturn, occurring
with opposite phases, are a well-known consequence of the so-called
``Great Inequality'', which is a near-resonance effect produced by
commensurability close to 2:5 between the mean motions of Jupiter and
Saturn \citep{lovett1895,michtchenko2001a,zink2020c}. It is further
worth noting that, although Saturn is significantly less massive than
Jupiter, the larger semi-major axis of Saturn results in a Hill radius
that is a factor of 1.22 times that of Jupiter.

According to Table~\ref{tab:mmr}, the 2:3 MMR ($\sim$6.8~AU) and 3:5
MMR ($\sim$7.3~AU) locations for Jupiter are almost identical to the
5:3 MMR and 3:2 MMR locations for Saturn, respectively. As seen in the
top panel of Figure~\ref{fig:sim78}, these strong resonance locations
demarcate a corresponding region of reduced scattering efficiency.

\subsection{Inclusion of the Ice Giants}
\label{ice}

The final scenario that was investigated (simulation 8) includes the
complete solar system giant planet suite of Jupiter, Saturn, Uranus,
and Neptune, all of which were considered to lie within their present
orbits. From an exoplanet perspective, true ice giant analogs are
challenging to detect given their relatively wide separations from the
host star \citep{kane2011d,wakeford2020b}. However, planet formation
models predict that Neptune-mass planets are relatively common
\citep{ida2005,emsenhuber2021c}, consistent with the findings of
microlensing surveys \citep{suzuki2018,zang2025}. Moreover, the
gravitational influence of Uranus and Neptune are particularly
interesting considering that their Hill radii are factors of 1.32 and
2.27 larger than Jupiter, respectively. Though the ice giants are far
outside the semi-major axis range of our primary simulations, they
serve as excellent examples of the continued effects giant planets can
exert over remaining solar system formation material. The ice giants,
especially Neptune, have played a key role in shaping the distribution
of material at the solar system outer edge and Kuiper Belt
\citep{levison2008a,dawson2012a}, including the capture of Kuiper Belt
Objects (KBOs) into Neptune resonance locations
\citep{malhotra1995b,hahn2005b,lykawka2007a}. Interaction with
planetesimals likely contributed to Neptune's migration
\citep{nesvorny2018c}, and the subsequent perturbing effects of
Neptune scattered material outward \citep{levison2003b}, producing a
population of high inclination KBOs \citep{gomes2003a}.

The results of simulation 8 are shown in the bottom panel of
Figure~\ref{fig:sim78}. Though the differences with simulation 7 are
subtle, the inclusion of the ice giants results in an overall decrease
in the scattering efficiency across the full semi-major axis range.
There are two primary causes of this decrease. First, similar to the
process described in Section~\ref{sat}, angular momentum is
transferred from Jupiter and Saturn to the ice giants, effectively
damping the eccentricities of Jupiter and Saturn. The eccentricity
ranges for simulation 8 are 0.0175--0.0579 and 0.0077--0.0611 for
Jupiter and Saturn, respectively, which are a slight decrease from the
eccentricity ranges provided in Section~\ref{sat} and partially
explains the decrease in simulation 8 scattering efficiency. Second,
since the simulations only include material interior to the orbit of
Saturn, the ice giants create a net outward flux of the material, thus
reducing the inward scattering potential of Jupiter and
Saturn. Although the nature of these planet-planet interactions are
sensitive to planetary architectures and initial conditions, it may be
the outward flux caused by the ice giants that dominates the results
for simulation 8. This could ideally be tested via future simulations
that extend the extent of the material semi-major axes out beyond the
orbit of Neptune.


\subsection{Comparative Scattering to the Inner Planets}
\label{comp}

The scenarios we have presented in the 8 simulations provide a variety
of orbital architectures and time frames over which to evaluate their
relative potential for scattering material interior to the snow line
and the inner solar system planets. To perform a comparison between
the simulation results, we calculated the total scattering efficiency
for each scattering threshold by integrating over all particle
locations. We excluded from the calculations those particles that lie
within three Hill radii ($\pm1$~AU) of Jupiter's semi-major axis to
account for planetesimal depletion due to Jupiter's ``feeding zone''
\citep{pollack1996,alibert2005b,raymond2017b}. The results of the
integrated scattering calculations are shown in
Table~\ref{tab:results} for all of the simulations, presented as a
percentage of the complete set of particles over the entire 3--8~AU
semi-major axis range.

\begin{deluxetable}{lrrrrr}
  \tablecolumns{6}
  \tablewidth{0pc}
  \tablecaption{\label{tab:results} Integrated scattering results.}
  \tablehead{
    \colhead{Sim} &
    \colhead{Mercury} &
    \colhead{Venus} &
    \colhead{Earth} &
    \colhead{Mars} &
    \colhead{Snow line}
  }
  \startdata
1 &  0.44 &  0.85 &  1.55 &  5.17 & 16.69 \\
2 & 28.55 & 36.45 & 43.67 & 57.57 & 76.54 \\
3 &  1.69 &  3.02 &  4.58 &  9.40 & 19.75 \\
4 & 38.74 & 48.43 & 56.61 & 70.42 & 87.13 \\
5 &  5.98 &  9.35 & 12.72 & 20.83 & 41.30 \\
6 & 61.75 & 72.49 & 79.58 & 89.21 & 96.38 \\
7 & 42.86 & 46.34 & 49.25 & 54.20 & 63.84 \\
8 & 39.92 & 42.95 & 45.40 & 49.74 & 58.84 \\
  \enddata
\end{deluxetable}

Referring to the simulation descriptions in Table~\ref{tab:sims},
simulations 1 and 2 consist of the the $10^5$~year integrations for
the Jupiter analog and $e = 0.23$ Jupiter cases, respectively. These
may be directly compared to simulations 3 and 4, which extend the
integration times to $10^6$~years. The differences between these
simulation runs are relatively minor when considering the snow line
threshold, but become increasingly important when evaluating the inner
solar system thresholds. For example, the $10^6$~year Jupiter analog
case (simulation 3) increases the scattering efficiency interior to
the snow line by a factor of 1.18 compared with the $10^5$~year case
(simulation 1), but the scattering efficiency interior to Earth's
orbit increases by a factor of 2.95. However, the $10^6$~year $e =
0.23$ Jupiter case (simulation 4) increases the scattering efficiency
interior to the snow line by a factor of 1.14 compared with the
$10^5$~year case (simulation 2), and the scattering efficiency
interior to Earth's orbit increases by a factor of 1.30. Though
calculations of scattering efficiency will obviously increase with
integration time, our results emphasize the particular importance of
this effect for low eccentricity cases, which can pose the risk of
significantly under-estimating the scattering rates.

Simulations 5 and 6 represent additional eccentricity scenarios for
Jupiter, specifically $e = 0.1$ and $e = 0.3$, respectively. As
described in Section~\ref{ecc}, the $e = 0.3$ scenario produces almost
complete scattering of material interior to the snow line for the full
range of semi-major axes explored, therefore representing the maximum
eccentricity in terms of optimized volatile scattering to the inner
system. In terms of scattering efficiency to the inner planets, the
effects of eccentricity become far more noticeable. For example, the
scattering efficiency to an Earth-crossing orbit is a factor of 17.4
times more for the $e = 0.3$ (simulation 6) case compared with the
Jupiter analog case (simulation 3). In general, moderate
eccentricities for Jupiter (0.2--0.3) result in an order of magnitude
increase in volatile delivery from beyond the snow line to the inner
planets.

The addition of Saturn in simulation 7 produces a scattering effect
that is more closely matched with the solar system architecture. As
noted in Section~\ref{sat}, the inclusion of Saturn produces angular
momentum exchanges that has the effect of periodic changes in
Jupiter's eccentricity that peak at $\sim$0.06. The total scattering
efficiency for the snow line threshold of 63.84\% is higher than that
of the $e = 0.1$ case (simulation 5), implying that the relatively
large Hill radius of Saturn is responsible for scattering the
remainder of the material. Overall, the effect of including Saturn on
the scattering efficiency interior to the snow line is equivalent to
an eccentric Jupiter scenario that lies in the range $e = 0.1$
(simulation 5) to $e = 0.23$ (simulation 4).

The addition of Uranus and Neptune in simulation 8 creates the full
suite of solar system giant planets as potential contributors to the
scattering efficiency. As described in Section~\ref{ice}, the
inclusion of the ice giants appears to slightly decrease the overall
scattering efficiency. Indeed, this is represented in
Table~\ref{tab:results}, where the scattering efficiency for the snow
line threshold is 5\% less than for the Jupiter/Saturn case
(simulation 7), showing that the ice giants have a very minor effect
on the scattering of material within the considered semi-major axis
range. Since the primary attribute explored in this work that
influences the scattering efficiency is the orbital eccentricity, the
scattering results are sensitive to architectures that draw angular
momentum away from Jupiter. Though their effect is minor, the
inclusion of Uranus and Neptune highlights the complexity of dynamical
interactions with multiple giant planets beyond the snow line that can
decrease the scattering efficiency for the primary perturber (in this
case, Jupiter). Other planetary architectures, such as those that
experience eccentricity excitation via planet-planet scattering or
resonance crossing events, may have a much more pronounced scattering
effect from even distant planets within such systems
\citep{chiang2002a,carrera2019b}.

A question remains as to how the scattering efficiency of material
beyond the snow line translates to potential volatile delivery for the
terrestrial planets. Material scattered into orbits that cross those
of the terrestrial planets does not guarantee collisions, and in fact
impact events are expected to comprise an extremely small fraction of
the overall scattered material
\citep{horner2008a,rickman2014c,strom2015}. Furthermore, although
impacts can contribute to the volatile inventory of a planet,
significant impacts can have the negative consequence of atmospheric
stripping
\citep{genda2003b,genda2005,newman1999,schlichting2015,kane2020d}. Planetary
impact probabilities for long-period comets were calculated by
\citet{zimbelman1984}, finding that the average impact probability per
periastron passage for Earth is $2.33\times10^{-9}$. For the other
terrestrial planets, the impact probabilities are 93\%, 170\%, and
12\% for Mercury, Venus, and Mars, respectively, relative to
Earth. However, as described in Section~\ref{discussion}, the
contribution of cometary impacts to volatile inventories are likely to
be appreciably smaller than those provided by planetesimal impacts
during the early stages of terrestrial planet formation.


\section{Discussion}
\label{discussion}

There are numerous important caveats to note regarding the work
presented in this paper. First, and most crucially, our simulations
only consider dynamical scattering of material after the gas phase
when planet formation has largely completed. Material present during
the gas phase experiences significant drag that governs its
distribution in addition to the gravitational influence of the forming
planets \citep{raymond2017b}. Furthermore, much of the material will
have been accreted or otherwise scattered by the planets during the
gas phase. Thus, it is recognized that the simulations described here
are applied to an inventory of remaining material that is
substantially lower that the mass of material present during the
planet formation process. Second, our simulations only consider
particles whose initial orbital inclination is coplanar with that of
Jupiter. Although material whose inclination is outside of the Jupiter
orbital plane will further contribute to volatile scattering, the
relative scattering rates described here remain valid for a comparison
between simulation models. Third, the integration duration of $10^6$
years at each semi-major axis location provides a modest improvement
to the $10^5$ year simulations when considering the snow line
threshold (see Section~\ref{comp}). However, the differences between
the $10^5$ and $10^6$ year simulations widen when considering
perihelion distances closer to the Sun, from which longer simulation
times may prove beneficial. Furthermore, increasing the simulation
integration times may narrow the regions of stability as more
gravitational perturbations are accounted for. Fourth, since our focus
is on the scattering caused by a Jupiter analog, we consider material
within the semi-major axis range 3--8~AU. The extension of the
simulations to farther distances would provide an improved assessment
of the scattering effects provided by planetary analogs to Saturn,
Uranus, and Neptune in supplying volatiles to the inner parts of
planetary systems.

The demographics of cold Jupiter systems is of increasing importance
as we seek to understand the architectures and evolution of inner
planetary systems
\citep[e.g.,][]{cumming2008,wittenmyer2011a,wittenmyer2020b,fulton2021,bonomo2023},
particularly in relation to volatile delivery mechanisms
\citep{raymond2004a,raymond2014d,ciesla2015b,raymond2017b,marov2018,venturini2020b}.
As described in Section~\ref{intro}, pebble drift theories provide a
potential pathway for volatile delivery from beyond the snow line
\citep{morbidelli2016a}. Such water delivery via pebbles is most
effective during terrestrial planet accretion
\citep{obrien2018,ida2019a}, and can depend strongly upon the radial
distribution of the protoplanetary disk and the structure of gaps that
may impede the inward flux of pebbles \citep{sato2016b}. Thus, the
extent of pebble delivery is sensitive to the assumed age and mass
distribution of the gas disk and the relative rates of planet
formation within the system. Similarly, the nature and volume of
volatile scattering that originates from giant planet perturbations
also depends on the distribution of volatiles throughout the
protoplanetary disk
\citep{pollack1996,lodders2003,arturdelavillarmois2019a,oberg2021b}. The
results presented here explore the scattering efficiency as a function
of semi-major axis with a flat distribution, and so may be scaled to
the known or inferred distribution of remaining material in the
post-gas phase within planetary systems.

Of special interest is the effect of giant planet formation on the
relative water content of the inner terrestrial planets
\citep{marty2012,morbidelli2012a,piani2020}. The inner solar system
planets are generally held to have formed via the accretion of
planetesimals and embryos within the protoplanetary disk
\citep{chambers1998b,kokubo2000a,morishima2010a,morbidelli2012a}. Since
Venus and Earth would have had similar feeding zones, it is expected
that they may have acquired similar amounts of water during formation
\citep{greenwood2018b,lykawka2024}, which has significant consequences
for the subsequent evolutionary history of Venus relative to an
initial dry model \citep{orourke2023}. Indeed, it is possible that
Venus could contain substantial amounts of water in its mantle, in
quantities comparable to that of Earth \citep{mccubbin2019a}. In
addition to the accretion processes, it is commonly inferred that
water was supplied via delivery from beyond the snow line, though the
relative contributions provided by accretion and delivery remain an
ongoing point of discussion
\citep{izidoro2013,raymond2017b,izidoro2022b}. The quantity of water
delivered can be highly sensitive to the planetary architecture and
the location of the snow line
\citep{ronco2014,mulders2015b,darriba2017}, and in some cases may have
a negligible \citep{quintana2014b} or negative \citep{sanchez2018}
effect on water delivery. However, a volatile-depleted inner solar
system may imply an increased dependency on delivery from beyond the
snow line \citep{albarede2009}. Similar to the studies of terrestrial
water gained via accretion, some models of water delivery suggest that
Venus may have received a comparable amount of volatiles to Earth, if
normalized to the unit mass of the planets \citep{marov2018}. These
relative amounts of delivered volatiles vary depending on atmospheric
surface pressures \citep{pham2011} and loss of the already present
volatiles due to ejection from impacts \citep{chyba1990a}. Our
simulations show that the full complement of giant planets (simulation
8) results in a comparable scattering of material into both Venus and
Earth-crossing orbits (see Table~\ref{tab:results}). However,
incorporating the relative impact probabilities derived by
\citet{zimbelman1984}, any water delivered from beyond the snow line
after the gas phase may have been substantially larger for Venus than
Earth. Given the current dry state of Venus, this may mean that water
delivery from the outer solar system constitutes a negligible amount
of the total volatile inventory of inner terrestrial planets, or that
Venus did indeed pass through a wet period with subsequent water loss
\citep{kasting1984a,way2016,kane2019d,kane2024b}. Indeed, the dramatic
loss of volatiles during the gas phase and the formation of the giant
planets will ensure that water delivery in the post-gas environment is
likely to represent a relatively minor fraction of the total volatile
delivery for the terrestrial planets.

Our results primarily investigate the dependence of remaining volatile
delivery on the eccentricity of the giant planets in the post-gas
phase. However, it is worth noting that early eccentricity of the
giant planets can in fact lead to dry inner planets by removing
volatile rich material from the inner protoplanetary disk
\citep{chambers2003b,raymond2004a}. Thus, the water inventory of the
solar system inner planets contains an imprint of not just the current
architecture, but is also sensitively dependent on the evolution of
the giant planets during their formation, emphasizing the need for a
correct diagnosis of the Grand Tack and Nice models of solar system
formation
\citep{gomes2005b,morbidelli2005,walsh2011c,raymond2014a,nesvorny2018c}.
Our simulations show that the role of Uranus and Neptune analogs
during post-formation scattering of volatiles is likely a combination
of their transferring of angular momentum from Jupiter and Saturn and
the outward flux of material. However, the extent that the latter
effect may be quantified from our work is limited by the 3--8~AU range
of our simulations.

As shown in Section~\ref{science}, there are currently 10 exoplanetary
systems with 2 or more known giant planets beyond the snow line, and
these systems likely harbor additional planets that have yet to be
unveiled. The systems contain exceptionally diverse architectures with
respect to the solar system, with stellar mass values of
0.42--1.09~$M_\odot$, planet mass values of 0.54--18.81~$M_J$, and
eccentricity values of 0.0--0.51. For the 8 non-circumbinary planetary
systems, the distances lie in the range 13.8--55.6~pcs, and the $V$
magnitudes in the range 5.0--12.4. These relatively bright stars thus
present numerous opportunities for follow-up observations, such as
extreme precision RV surveys to discover additional planets and refine
orbits \citep{kane2009c,fischer2016,pepe2021,kane2024e}, and
constraining planet formation and structure through the study of
stellar composition \citep{hinkel2019,pignatari2023a}. Furthermore,
direct imaging for the nearest stars may enable further
characterization of the giant planets within those systems
\citep{kane2013c,kopparapu2018,stark2024}. Of particular interest is
47~Uma, a star known to harbor a total of 3 planets
\citep{rosenthal2021}. The distance (13.8~pcs) and snow line (3.03~AU)
for 47~Uma yields an angular separation of the snow line of
0.22~arcsec. Combined with the brightness ($V = 5.0$), the above
described features make 47~UMa an attractive target for direct imaging
observations \citep{li2021a,turnbull2021}.


\section{Conclusions}
\label{conclusions}

Planetary habitability is intrinsically linked to the ability of a
terrestrial planet to sustain the presence of surface liquid water
over long time scales. This sustainability is, in turn, dependent on
the initial volatile inventory of the planet that was acquired during
and after formation. For the solar system, the role of Jupiter in the
delivery of Earth's volatiles has been a matter of substantial
research. As the dominant planetary mass within the solar system,
Jupiter has greatly influenced the sculpting of the solar system
architecture and the distribution and scattering of volatiles that
predominantly lie beyond the system snow line. Likewise, the formation
and evolution of other planetary system architectures, and their
influence on the volatile inventory of the terrestrial planets, is a
growing area of astrobiological interest. Given the relative scarcity
of cold giant planets in other systems, a thorough assessment of our
giant planets impact on Earth's evolutionary history is one of
critical importance for adjudicating the likelihood of similar
habitable conditions elsewhere.

Our dynamical simulations provide a quantitative framework from which
to calculate the relative scattering potential for a variety of
architecture models beyond the gas phase of planet formation.  The
case of a Jupiter analog with an increased eccentricity can result in
a significant increase in the scattering efficiency interior to the
snow line. Interestingly, a giant planet pair, such as Jupiter and
Saturn, can produce a scattering efficiency that is equivalent to a
single Jupiter analog with an eccentricity in the range 0.2--0.3. A
consequence of this result is that even exoplanetary systems with a
single giant planet beyond the snow line may enable additional
volatile delivery to inner terrestrial planets after their formation
and insofar as sufficient material remains to be scattered. On the
other hand, planets at the very outer edge of the system, such as
Uranus and Neptune, can have a damping effect on the orbital
eccentricities of inner giant planets, slightly decreasing the
scattering of available material. The diversity of exoplanetary
architectures provide a means of testing the various scattering models
through atmospheric characterization of terrestrial planets, and
gradually revealing the true role of giant planets in the evolution of
planetary habitability.


\section*{Acknowledgements}

The authors would like to thank Sean Raymond for his valuable feedback
on the manuscript. This research has made use of the Habitable Zone
Gallery at hzgallery.org. The results reported herein benefited from
collaborations and/or information exchange within NASA's Nexus for
Exoplanet System Science (NExSS) research coordination network
sponsored by NASA's Science Mission Directorate.


\software{Mercury \citep{chambers1999}}




\end{document}